\documentclass[aps,epsfigure,twocolumn,superscriptaddress,showkeys,nofootinbib]{revtex4-1}
\usepackage[colorlinks=true,linkcolor=blue,urlcolor=blue,citecolor=blue,pdfusetitle]{hyperref}
\usepackage[utf8]{inputenc}
\usepackage[english]{babel}
\usepackage{amsmath}
\usepackage{chemfig}
\usepackage{graphicx, float, epstopdf}
\usepackage{blindtext}
\usepackage{amsfonts}
\usepackage{bbm}
\usepackage{amssymb}
\usepackage{enumerate}
\usepackage{color}
\usepackage{latexsym}
\usepackage{physics}
\usepackage{times,txfonts}
\usepackage[normalem]{ulem}
\usepackage{subfigure}
\usepackage{comment}
\usepackage{booktabs}

\usepackage{amsmath}
\usepackage{qcircuit}

\begin{document}

\title{Exploiting the Variational Quantum Eigensolver for Determining Ground State Energy of Protocatechuic Acid}

\author{Gleydson Fernandes de Jesus}
\email{gleydson.jesus@fieb.org.br}
\affiliation{QuIIN - Quantum Industrial Innovation, EMBRAPII CIMATEC Competence Center in Quantum Technologies, SENAI CIMATEC, Av. Orlando Gomes, Salvador, BA 1845, Brazil.}
\affiliation{Latin America Quantum Computing Center, SENAI CIMATEC, Av. Orlando Gomes, 1845, Salvador, BA, Brazil CEP 41850-010.}

\author{Erico Souza Teixeira}
\email{est@cesar.org.br}
\affiliation{CESAR, Recife, Brazil}
\affiliation{$|QATS\rangle$ - Quantum Application in Technology and Software Research Group.}

\author{Lucas Queiroz Galvão}
\email{lucas.queiroz@fbter.org.br}
\affiliation{QuIIN - Quantum Industrial Innovation, EMBRAPII CIMATEC Competence Center in Quantum Technologies, SENAI CIMATEC, Av. Orlando Gomes, Salvador, BA 1845, Brazil.}
\affiliation{Latin America Quantum Computing Center, SENAI CIMATEC, Av. Orlando Gomes, 1845, Salvador, BA, Brazil CEP 41850-010.}

\author{Maria Heloísa Fraga da Silva}
\email{maria.fraga@fbter.org.br}
\affiliation{QuIIN - Quantum Industrial Innovation, EMBRAPII CIMATEC Competence Center in Quantum Technologies, SENAI CIMATEC, Av. Orlando Gomes, Salvador, BA 1845, Brazil.}
\affiliation{Latin America Quantum Computing Center, SENAI CIMATEC, Av. Orlando Gomes, 1845, Salvador, BA, Brazil CEP 41850-010.}
\affiliation{Grupo de Informação Quântica e Física Estatística, Centro de Ciências Exatas e das Tecnologias, Universidade Federal do Oeste da Bahia---Campus Reitor Edgard Santos. Rua Bertioga, 892, Morada Nobre I, 47810-059 Barreiras, Bahia, Brazil;}

\author{Mauro Queiroz Nooblath Neto}
\email{mauro.neto@fieb.org.br}
\affiliation{QuIIN - Quantum Industrial Innovation, EMBRAPII CIMATEC Competence Center in Quantum Technologies, SENAI CIMATEC, Av. Orlando Gomes, Salvador, BA 1845, Brazil.}
\affiliation{Latin America Quantum Computing Center, SENAI CIMATEC, Av. Orlando Gomes, 1845, Salvador, BA, Brazil CEP 41850-010.}

\author{Bruno Oziel Fernandez}
\email{bruno.fernandez@fieb.org.br}
\affiliation{Latin America Quantum Computing Center, SENAI CIMATEC, Av. Orlando Gomes, 1845, Salvador, BA, Brazil CEP 41850-010.}

\author{Clebson dos Santos Cruz}
\email{clebson.cruz@ufob.edu.br}
\affiliation{Grupo de Informação Quântica e Física Estatística, Centro de Ciências Exatas e das Tecnologias, Universidade Federal do Oeste da Bahia---Campus Reitor Edgard Santos. Rua Bertioga, 892, Morada Nobre I, 47810-059 Barreiras, Bahia, Brazil;}

\begin{abstract}

The Variational Quantum Eigensolver (VQE) is a promising hybrid algorithm, utilizing both quantum and classical computers to obtain the ground state energy of molecules. In this context, this study applies VQE to investigate the ground state of protocatechuic acid, analyzing its performance with various \textit{Ansätze} and active spaces. Subsequently, all VQE results were compared to those obtained with the Hartree-Fock (HF) method. The results demonstrate that \textit{Ansätze}, like unitary coupled-cluster singles and doubles (UCCSD) and variations of Hardware-Efficient, generally achieve accuracy close to that of HF. {Furthermore, the increase in active space has led to the models becoming more difficult to converge to values with a greater distance from the correct energy}. In summary, the findings of this study reinforce the use of VQE  as a powerful tool for analyzing molecular ground state energies. Finally, the results underscore the critical importance of \textit{Ansatz} selection and active space size in VQE performance, providing valuable insights into its potential and limitations. 
\end{abstract}


\maketitle

\section{Introduction}

The challenge of predicting molecular energies through the Schrödinger equation is compounded by the lack of exact solutions for multi-electron systems, necessitating the development of various electronic structure methods. These methods, while successful, often face significant computational costs that escalate with system size and complexity, particularly for high-accuracy approaches that may scale exponentially, demanding extensive memory and computational time \cite{khalid_reggab_2024}. Recent advancements aim to strike a balance between accuracy and computational efficiency, with new algorithms and approximations being explored to achieve chemical precision at a more manageable cost \cite{shuiping_zhao__2024,sayan_maity__2023, Manzhos_2020}. This progress is crucial for enhancing the synergy between theoretical predictions and experimental validations, ultimately facilitating more effective applications in materials science and chemistry. However, the ongoing need for improved methods highlights the limitations of current approaches and the importance of continued research in this area \cite{khalid_reggab_2024}.

The advancement of quantum computing is poised to revolutionize problem-solving in chemistry, particularly through the development of innovative quantum algorithms. Research indicates that while current quantum algorithms for simulating electronic ground states may be slower than classical methods like Hartree-Fock (HF), they offer potential advantages in accuracy and may avoid some of the exponential scalability issues faced by classical approaches, particularly as quantum hardware improves \cite{werner_dobrautz__2024}. For instance, early studies suggest that quantum algorithms could eventually tackle complex chemical challenges that are difficult for classical computers, showcasing significant potential in this field \cite{rajesh_kumar_tiwari__2024, javier_robledomoreno__2024}. Furthermore, the exploration of quantum computing's capabilities in chemistry is gaining momentum, with numerous studies highlighting ongoing efforts to refine these algorithms and enhance their efficiency \cite{cao2019quantum, McArdle_2020, Blunt2022Perspective, Choy2023Molecular, Evangelista2023}. Despite the current limitations in speed and hardware, the potential accuracy and scalability advantages of quantum algorithms suggest a promising future for their application in solving intricate chemical problems.

Quantum Phase Estimation (QPE) is indeed a pivotal algorithm in quantum computing, particularly for simulating molecular systems and determining their energetic properties \cite{O’Brien2018Quantum, O’Brien2019Calculating, Bauman2020Toward}. It allows for precise energy estimates of quantum states, which are crucial for understanding chemical and physical characteristics \cite{mohammadbagherpoor2019improved}. However, practical limitations significantly hinder its implementation. Firstly, QPE demands a large number of qubits and quantum operations, which scale with the required precision and the complexity of the system, challenging the capabilities of current quantum hardware \cite{dimitris_ntalaperas__2024}. Additionally, the algorithm's performance is adversely affected by noise and decoherence, which can lead to inaccuracies in energy estimations \cite{charles_woodrum__2024}, particularly due to the long coherence times and high precision required. Furthermore, the necessity for well-defined unitary operators complicates implementation in real-world scenarios, as these operators must be accurately constructed and maintained, often requiring deep quantum circuits \cite{kumar2022quantum, prokopenya2015simulation}. These factors collectively underscore the challenges faced by QPE in practical applications, necessitating ongoing research to enhance its robustness and efficiency \cite{PhysRevA.109.032606}.

The Variational Quantum Eigensolver (VQE) \cite{tilly2022variational} has emerged as a promising approach to tackle scalability challenges in quantum computing, particularly in quantum chemistry applications. By integrating quantum variational principles with classical optimization, VQE demonstrates resilience to errors from decoherence and noise, although this resilience has its limits. Recent studies highlight VQE's capability to accurately compute molecular energies for simple systems like $H_{2}$ \cite{singh2024ground}, and $BeH_{2}$ \cite{kandala2017hardware}, showcasing its potential for more complex molecules. Researchers are actively investigating enhancements to VQE's efficiency and scalability, aiming to extend its applicability to larger systems \cite{lim2024fragment}. In this context, we propose using VQE to find the ground state of protocatechuic acid (PCA), a chemical compound derived from benzoic acid with significant antioxidant properties for the pharmaceutical and food industries \cite{kakkar2014review}. This ongoing exploration underscores VQE's critical role in advancing quantum chemistry and its practical applications.

The remainder of the article is organized as follows. In Sec. \ref{sec:methods}, we describe the methodology. In Sec. \ref{sec:result}, we discuss our simulations. Finally, we conclude our work in Sec. \ref{sec:result}, and Appendix \ref{apend:ansatze} shows the details of the variational circuit used.

\section{Material and Methods}\label{sec:methods}

\subsection{Protocatechuic Acid}

\begin{figure}[h]
\centering
\includegraphics[width = 4cm]{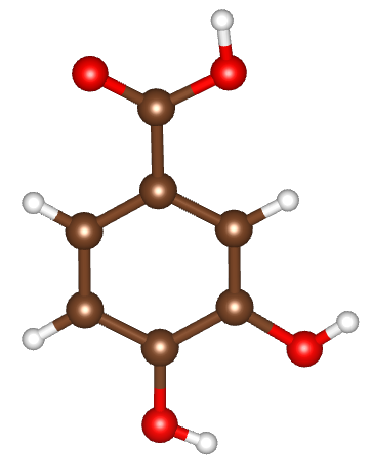}
\caption{Chemical structure of protocatechuic acid ($C_{7}H_{6}O_{4}$), also known as 3,4-dihydroxybenzoic acid. The molecule consists of a benzene ring substituted with two hydroxyl groups (-OH) at the 3rd and 4th positions and a carboxyl group (-COOH) at the 1st position. This arrangement classifies it as a dihydroxybenzoic acid, contributing to its antioxidant properties commonly found in various plants.  C: brown, O: red, H: white.}
\label{fig:molecule}
\end{figure}

In this study, we performed a quantum algorithm to find the ground state of protocatechuic acid (PCA), derived from benzoic acid and characterized by the molecular formula $C_{7}H_{6}O_{4}$ with chemical name 3,4-dihydroxybenzoic acid \cite{acid}, whose flat structural formula is shown in Figure \ref{fig:molecule} \cite{Momma}.

As a phenolic acid, PCA has an aromatic ring and hydroxyl groups in specific positions that give it important antioxidant properties for the pharmaceutical and food industries \cite{kakkar2014review}. In medicine, recent studies suggest that protocatechuic acid has chemopreventive activities against cancer, anti-inflammatory, cardiopreventive, neuroprotective, antibacterial, antiviral, and anti-aging properties \cite{song2020new, zhao2021cholesterol, nam2018protocatechuic}.
 
In terms of occurrence, PCA is an active constituent of more than 500 plants, including the \textit{Euterpe oleracea} \cite{oliveira2022structural}. This typical Amazonian palm serves as the raw material for the extraction of açaí oil, a compound rich in protocatechuic acid (630$\pm$36 mg/kg) and widely consumed in Brazil \cite{pacheco2008chemical}.

\subsection{Basis Set}
We use the least computationally demanding basis, STO-3G, to minimize the number of qubits needed for computation. In this basis, each hydrogen (H) atom is represented by a single 1s orbital, while each carbon (C) and oxygen (O) atom is represented by a set of orbitals: 1s, 2s, 2px, 2py, and 2pz \cite{jensen2017introduction}. Therefore, handling the entire molecule 
$C_{7}H_{6}O_{4}$ requires considering 61 orbitals or, equivalently, 122 spin orbitals, which would require 122 qubits.

\subsection{Active-space}
Modeling electronic structure problems with quantum computing is a complex task, primarily due to the linear scaling of the necessary qubits with the number of spin orbitals, which is twice the number of spatial orbitals in the basis set. This challenge is further compounded by the need to use small basis sets and the active space approximation \cite{jensen2017introduction}, which involves selecting a subset of Molecular Orbitals (MOs) for the calculation \cite{von2021quantum}. The classification of molecular orbitals into core, active, and external orbitals adds another layer of complexity \cite{bensberg2023corresponding}, with core orbitals typically fully occupied by two electrons, active orbitals potentially occupied by zero, one, or two electrons, and external orbitals generally not included in the calculation \cite{pennylanedoc}. {Since simulating the entire molecule $C_{7}H_{6}O_{4}$ requires 122 qubits, a number that exceeds the capabilities of our simulator (which can handle up to 35 qubits), we have reduced the size of the system through active-space selection.}

In our investigation, we are focusing on a crucial aspect of the acid group's conjugation with the aromatic ring, leading to internal resonance. This focus is driven by the fact that most of the electronic behavior of aromatic molecules is described by the $\pi$ orbitals of carbon atoms, and the two highest occupied orbitals in benzene are degenerate. Our initial focus is on a 4-electron, 4-orbital active space corresponding to an 8-qubit case \cite{martinez2022molecular}. We are also exploring larger active spaces within the capacity of our hardware, considering 6 and 8 orbitals for the active space, which correspond to 12 and 16 spin orbitals, represented by 12 and 16 qubits, respectively \cite{atkins2011molecular}.

\subsection{Reference Energy}
\begin{figure*}[ht]
    \centering
    \includegraphics[width = \textwidth]{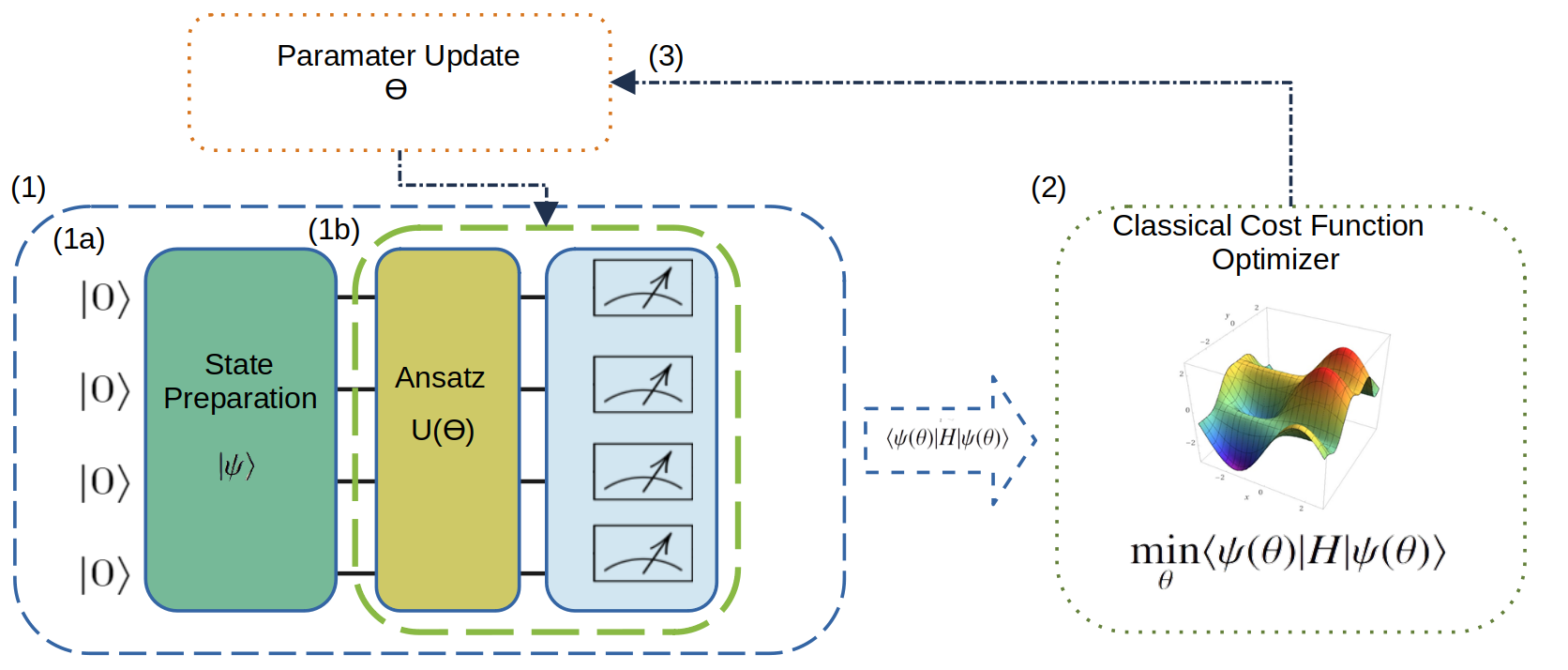}
    \caption{(1a) Qubits are initialized using the Jordan-Wigner transformation, mapping Molecular Orbitals (MOs) onto qubits to represent the initial quantum state $|\psi\rangle$. (1b) A parameterized quantum circuit, the ansatz $U(\theta)$, is applied to the initial state to generate $|\psi(\theta)\rangle$, which encodes a possible solution. The system’s Hamiltonian $\hat{\mathcal{H}}$ acts as the cost function, and the quantum circuit is measured to compute the expected value $\langle \psi(\theta)|\hat{\mathcal{H}}|\psi(\theta)\rangle$. (3) A classical optimization loop iteratively updates the parameters $\theta$ to minimize the cost function, adjusting the ansatz and recalculating until convergence. The final optimized value represents the ground state energy of the quantum system.}
    \label{fig: fig1}
\end{figure*}

The energy applied as a reference to evaluate the precision of quantum methods was determined using the molecular electronic Hamiltonian calculated using the Hartree-Fock method \cite{szabo1996modern}. This approach treats each electron as an independent particle influenced by the Coulomb potential from the nuclei and an averaged field from other electrons, allowing for a simplified yet effective representation of electron interactions. The exact diagonalization technique \cite{weisse2008exact} is then employed to derive energy eigenvalues from this Hamiltonian, where the system is represented in matrix form and diagonalized to obtain precise energy levels. Research indicates that while the Hartree-Fock method provides a foundational understanding, it may only capture some correlation effects present in more complex systems, potentially leading to inaccuracies in energy predictions that could hinder scientific progress.

\subsection{Variational Quantum Eigensolver}

Variational Quantum Algorithms (VQA) are widely applied in quantum computing due to their remarkable flexibility in tackling various categories of problems, especially in optimization and quantum machine learning \cite{Cerezo_2021}. These algorithms are designed to minimize or maximize a cost function $C(\theta)$ based on the foundations of the variational principle of quantum mechanics \cite{Cerezo_2021, bittel2021training}. One of the most commonly used VQA is the Variational Quantum Eigensolver (VQE) \cite{cerezo2022variational}, specifically targets finding the ground state energy of a quantum system.

In quantum mechanics, the Hamiltonian 
$\hat{\mathcal{H}}$ determines a system's total energy, with its eigenvalues corresponding to the energy levels. The ground state energy  $E_{0}$ is the lowest eigenvalue of $\hat{\mathcal{H}}$
and corresponds to the most stable state of the system. Mathematically, if $|\psi_0\rangle$ is the ground state, then  $\hat{\mathcal{H}} | {\psi_0} \rangle = E_0 |\psi_0 \rangle$	is the lowest eigenvalue. The variational principle ensures that the expectation value of $\hat{\mathcal{H}}$ with respect to any trial wavefunction provides an upper bound to $E_{0}$, the lowest eigenvalue of $\hat{\mathcal{H}}$:

\begin{equation}
    E_{0} = \frac{\langle \psi_0 | \hat{\mathcal{H}} | \psi_0 \rangle}{\langle \psi_0 | \psi_0 \rangle}.
\label{eq:vqe_cq}
\end{equation}
In practice, VQE optimizes a trial wavefunction, $|\psi (\theta)\rangle $, parameterized by $\theta$ to minimize the expectation value: 

\begin{equation}
    E_{0} \leq \frac{\langle \psi(\theta)  | \hat{\mathcal{H}} | \psi (\theta) \rangle}{\langle \psi (\theta) | \psi(\theta) \rangle}.
\label{eq:vqe_cq1}
\end{equation}
Through this optimization process, VQE approximates the lowest possible value of this expectation value, providing an estimate of the ground state energy. That algorithm can be described as follows:

\begin{enumerate}
    \item \textbf{Initialize the quantum state  $|\psi\rangle$:} the initial state that qubits are put before VQE algorithm start (Figure \ref{fig: fig1} 1a). We use the Jordan-Wigner transformation to map Molecular Orbitals (MOs) onto qubits;

    \item \textbf{Define the Hamiltonian H: } represent the energy operator of the quantum system as the Cost function. At Figure \ref{fig: fig1}, it is part of the measurement circuit at (Figure \ref{fig: fig1} 1b);
   
    \item  \textbf{Construct an ansatz $U(\theta)$:} represent the search space of possible solutions as a parameterized circuit ansatz (Figure \ref{fig: fig1} in 1b). Apply this operator at $|\psi\rangle$ generate $|\psi(\theta)\rangle$ as represented in Equation \ref{eq:psitheta};
    \begin{equation} \label{eq:psitheta}
    |\psi(\theta)\rangle = U(\theta)|\psi\rangle \end{equation}
    
    \item \textbf{Initialize Parameters:} randomly or based on prior knowledge about the problem, associated values to the ansatz's parameters as illustrated in Figure \ref{fig: fig1} in (3);

    \item \textbf{Measure:} compute the expected value of the cost function $C(\theta) = \langle \psi(\theta)|\hat{\mathcal{H}}|\psi(\theta)\rangle$. This process involves executing the quantum circuit corresponding to the \textit{Ansatz} and subsequently measuring the qubits to determine the expected value of each term in the $\hat{\mathcal{H}}$ (Figure  in \ref{fig: fig1} 1b);
    
    \item \textbf{Optimize Parameters:} tuning \textit{Ansatz} parameters ($\theta$) using classical optimization algorithms to minimize the Hamiltonian's expected value;

    \item \textbf {Convergence:} repeat the fifth and sixth steps until we achieve the optimal solution that minimizes $C(\theta$). From a physical point of view, this value represents the energy of the ground state of a given quantum system.
    
\end{enumerate}

\begin{figure*}[ht]
     \centering
     \includegraphics[scale = 0.75]{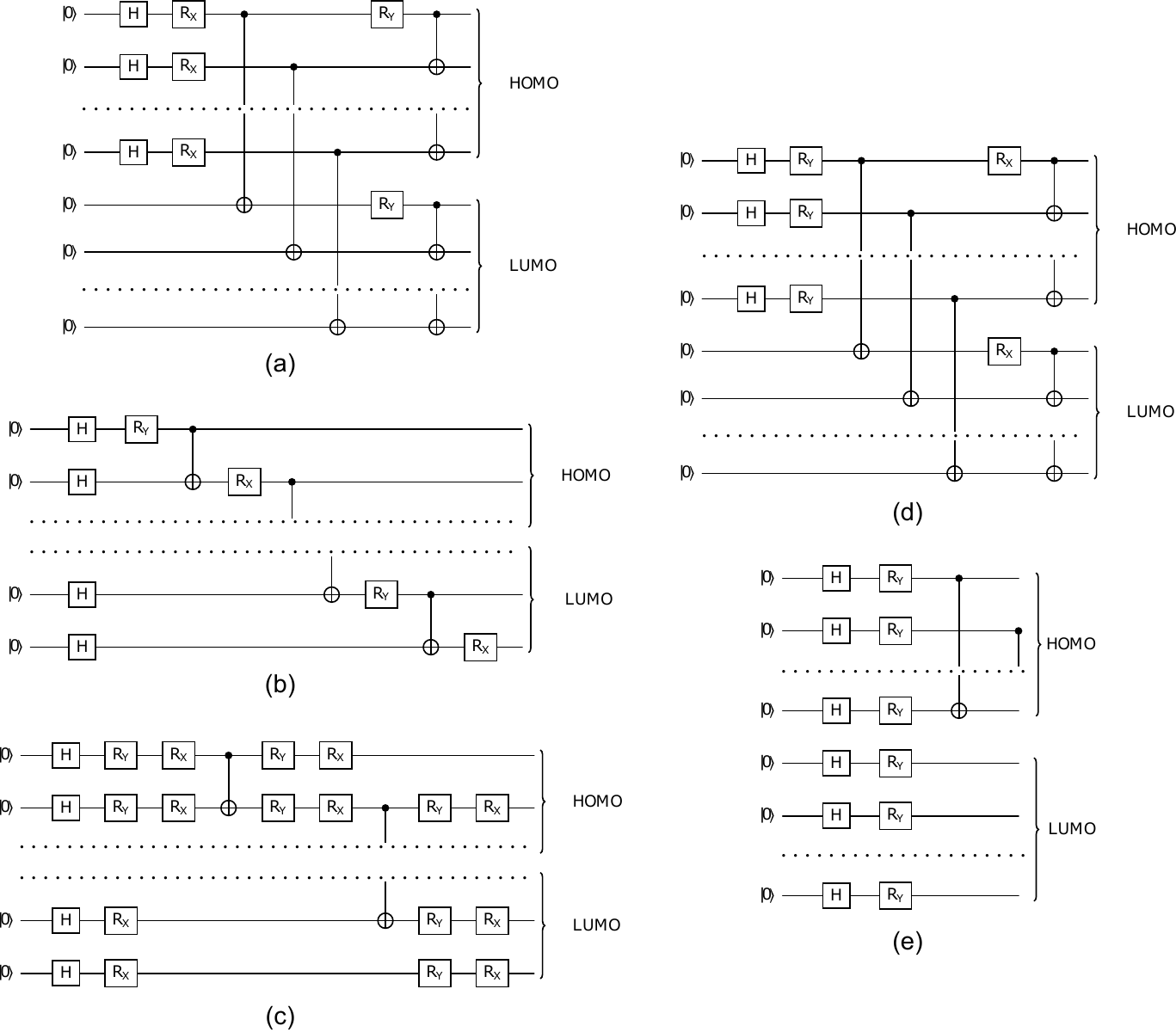}
     \caption{Hardware-efficient quantum circuits designed to simulate the protocatechuic acid molecule. Each subfigure represents a different circuit version with varying parameters and gate counts. (a) Circuit v1 includes $N_{qubits}$ parameters and $5N_{qubits}/2$ gates, balancing simplicity and computational power. (b) Circuit v2 similarly uses $N_{qubits}$ parameters and $5N_{qubits}/2$ gates. (c) Circuit v3 features more complexity with $2(3N_{qubits} - 2)$ parameters and $15N_{qubits}/2$ gates, making it suitable for more detailed simulations. (d) Circuit v4 employs $N_{qubits}$ parameters, offering a more minimalistic approach, while (e) Circuit v5 uses $N_{qubits}$ parameters and $9N_{qubits}/4$ gates, providing a moderate level of gate efficiency for the simulation.}
     \label{fig:ansatz} 
\end{figure*} 

\subsection{Anzätze}

The choice of ansatz in variational algorithms is pivotal, as it directly impacts the algorithm's ability to approximate quantum states effectively. A well-chosen ansatz can enhance computational efficiency while maintaining accuracy, which is essential for capturing the intricate correlations within quantum systems \cite{cao2019quantum}. Research indicates that simplistic ansatz forms may overlook critical interactions \cite{lorenzo_leone__2024}, leading to significant discrepancies in results, while overly complex ansatz can result in prohibitive computational demands, thus hindering practical applications. For instance, studies highlight the importance of tailoring the ansatz to the specific characteristics of the quantum system under investigation, as this can optimize performance and resource utilization. Moreover, the balance between complexity and efficiency is crucial; an ideal ansatz should be flexible enough to adapt to various quantum states without incurring excessive computational costs. Therefore, careful selection and design of the ansatz are fundamental to the success of variational algorithms in quantum computing.

For the simulations carried out in this paper, five HE ansatz circuits were considered (Figure \ref{fig:ansatz}). The first three are ansatz (v1, v2, v3) used in reference \cite{sennane2023calculating} to calculate the ground state of the benzene molecule. The v4 was proposed considering changes in the rotations of the ansatz V1. The v5 is an ansatz proposed in this work considering the coupling of qubits in the HOMO layer.

The ansatze 1 and 4 can be understood in the context of chemical interpretation. In these ansatze, the first $N_{qubits}/2$ qubits are used to represent the occupied orbitals, and the Hadamard gate is applied to create an electron on each. The first set of CNOT entanglements in both ansatze creates an occupied-virtual mixing, and the second set creates a spin mixing. The difference lies in the order of application of the RX and RY gates. The ansatz 4 follows the order of an RY gate followed by an RX gate, which is the same as a Hadamard gate. In contrast, ansatz 2 creates electrons in all the orbitals with less entanglement (CNOT gates) but with similar freedom to adjust it by a similar number of rotation gates. Ansatz 3 also creates electrons in all the orbitals, but when compared with ansatz 2, it gives more freedom to adjust the entanglement, i.e., the occupancy of each orbital, by increasing the number of rotation gates. At least, ansatz 5 is the simplest one, creating electrons in all orbitals, but fewer entanglement and freedom than the previous one.

\subsection{Simulation Environment (PennyLane)}

All the simulations were performed in the PennyLane \cite{Pennylane} quantum computing simulation environment, which was developed by the quantum computing company Xanadu \cite{Xanadu}. 

PennyLane is a library for simulating quantum computing algorithms, such as quantum chemistry algorithms and quantum machine learning algorithms, and it has resources for running these algorithms on different hardware \cite{PennylanePlugins}. 
The calculation of the molecular Hamiltonians was performed sequentially in a single-core, due to PennyLane's lack of standard parallelization for this task. The calculation of the reference energy was performed on an Intel Xeon Platinum 8260L processor, and the calculation of energies using VQE was performed on an Nvidia Tesla V100 GPU.

\section{Result and Discussion}\label{sec:result}
\begin{figure}[H]
    \centering
    \subfigure[]{\includegraphics[width = \columnwidth]{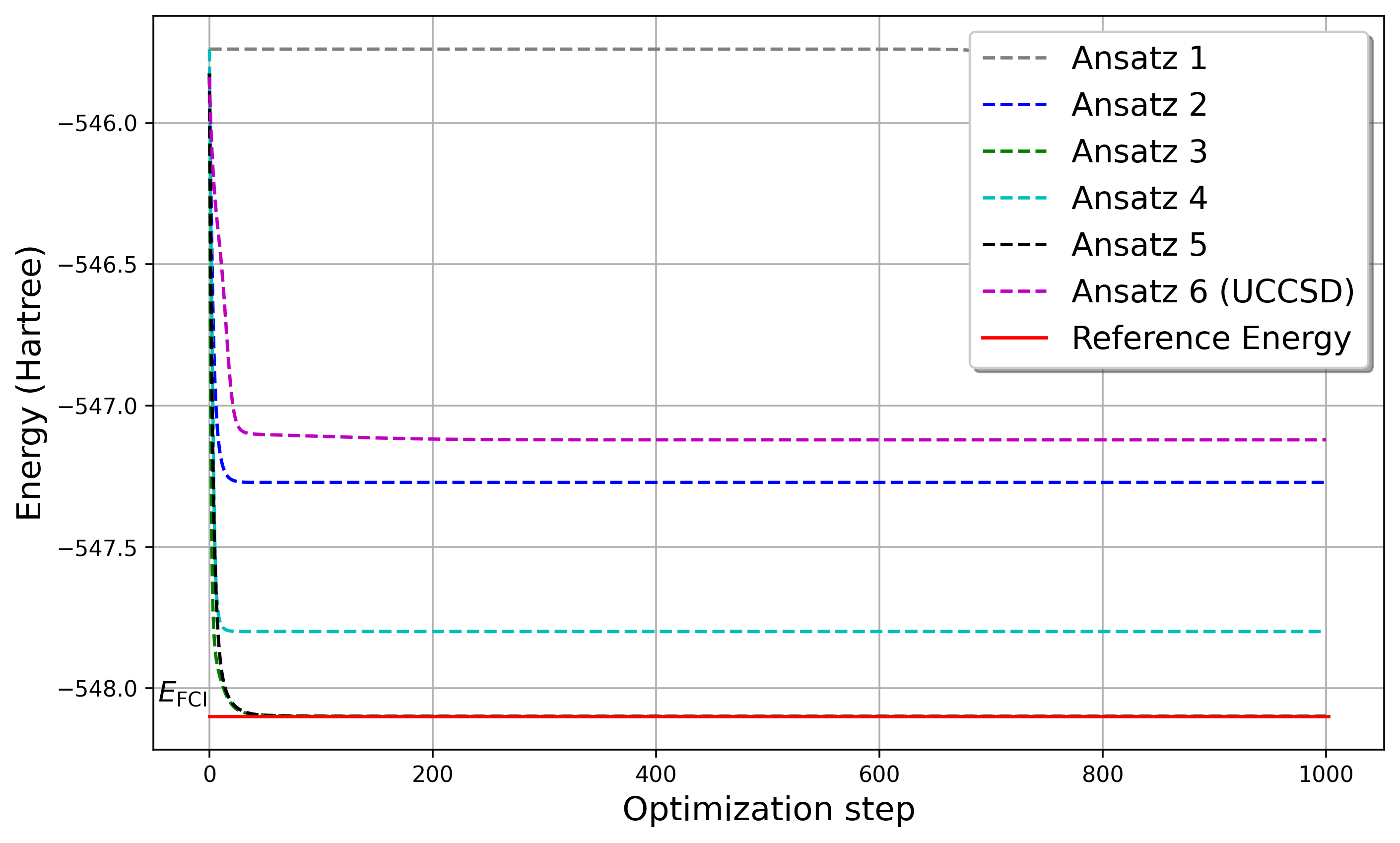}}
    \subfigure[]{\includegraphics[width = \columnwidth]{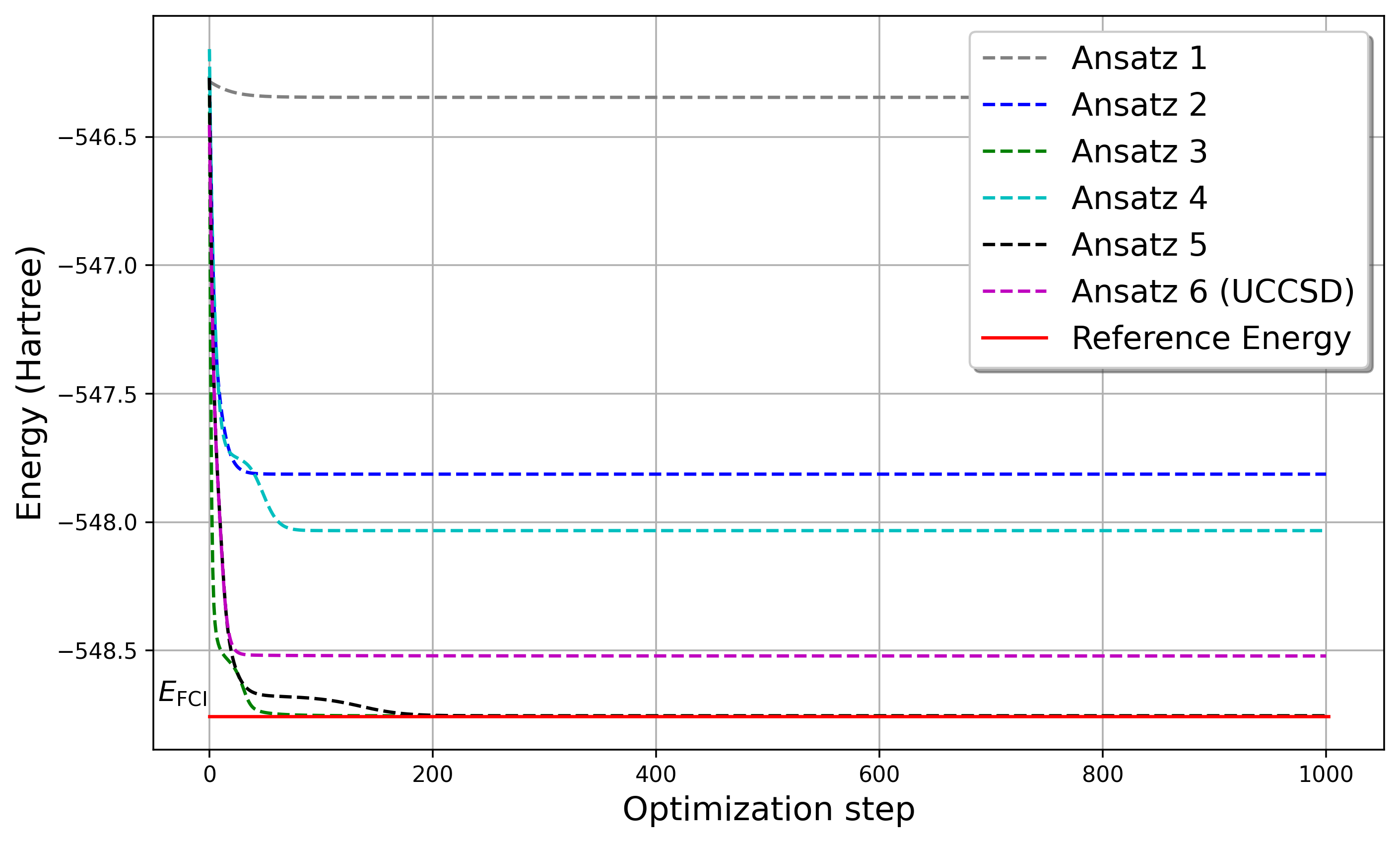}}
    \subfigure[]{\includegraphics[width = \columnwidth]{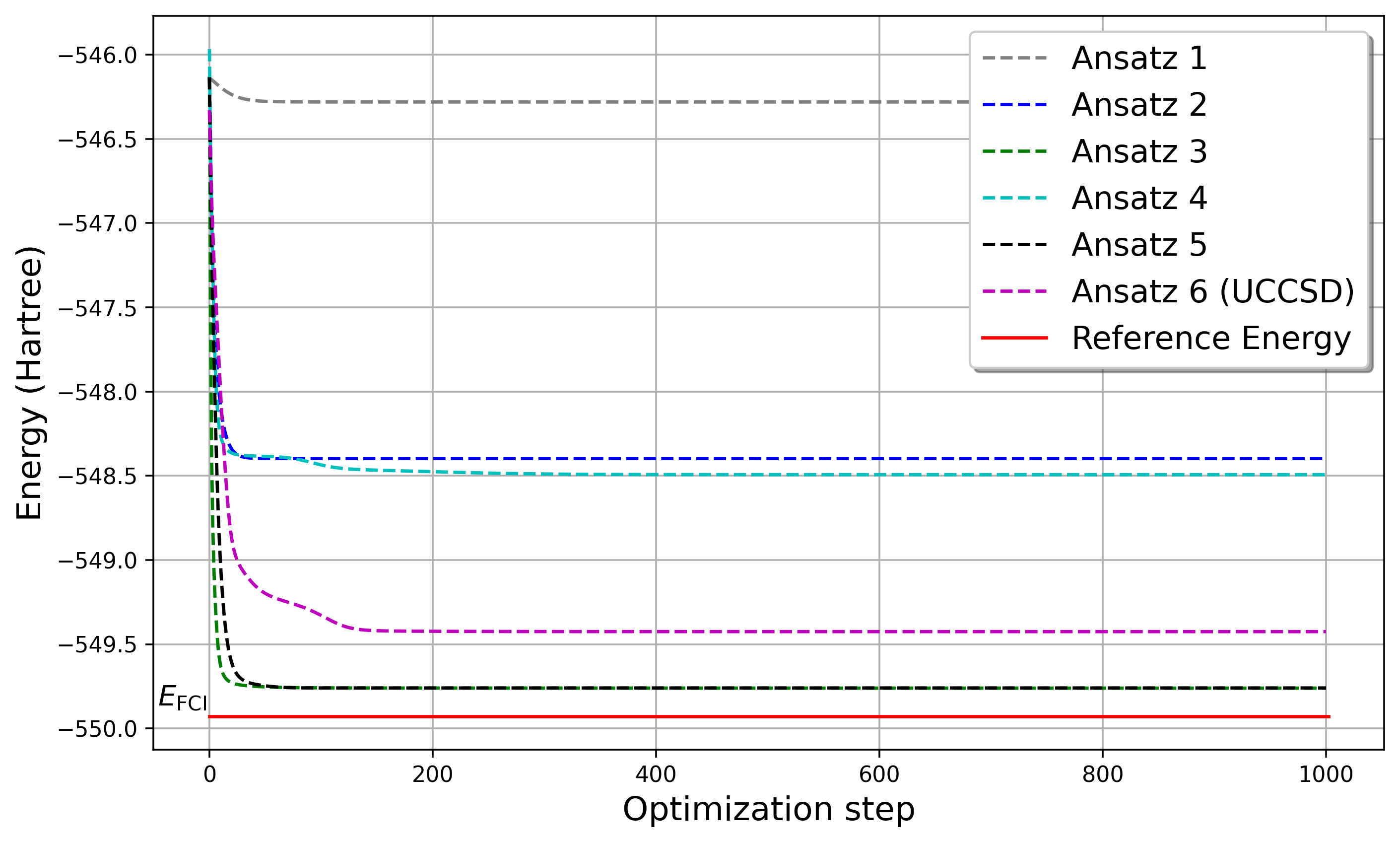}}
    \caption{Optimization process of protocatechuic acid simulation with varying numbers of molecular orbitals (MOs) in the active space: (a) 4 MOs, (b) 6 MOs, and (c) 8 MOs. The x-axis represents the number of optimization steps, while the y-axis displays the corresponding energy in Hartrees at each step. The dotted lines illustrate the energy convergence for each of the 5 ansätze utilized in the simulations, indicating their respective trajectories toward energy minimization. The red horizontal line denotes the reference energy, providing a benchmark for evaluating the accuracy and performance of the different ansätze across all active space configurations during the optimization process.}
    \label{fig:vqe4orbitals}
\end{figure}

The results obtained from the VQE simulations using six different ansatzes and reducing the active space to four, six, and eight molecular orbitals are illustrated in Figure \ref{fig:vqe4orbitals}. Additionally, the reference energy, determined by diagonalizing the molecule’s Hamiltonian for each configuration, is presented for comparison.

Ansatzes 3 and 5 consistently demonstrated superior performance across all cases. For the four- and six-orbital active spaces, both ansatzes achieved chemical precision, with errors less than $1.6 \times 10^{-3}$ Ha \cite{kandala2017hardware}. However, for the eight-orbital active space, the errors increased to approximately $1.68 \times 10^{-1}$ Ha, exceeding chemical precision. It is noteworthy that although ansatzes 3 and 5 have distinct structures, they converged to identical results. While ansatz 3 employs a greater number of gates and parameters, ansatz 5 utilizes fewer. Interestingly, in the six- and eight-orbital cases, ansatz 5 converged in fewer iterations, highlighting its efficiency.
Ansatzes 2 and 4 showed moderate performance, while ansatz 1 consistently underperformed, yielding results far from the expected values and remaining close to the system's initial energy.

The differences in performance can be attributed to the excitation possibilities explored by each ansatz. Ansatzes 1 and 4 limit excitations by applying Hadamard gates solely to the HOMOs and restricting entanglement to the HOMO and LUMO orbitals (Figures \ref{fig:ansatz} (a) and \ref{fig:ansatz} (b)). 

In contrast, ansatzes 2, 3, and 5 apply Hadamard gates to all qubits, enabling entanglements between HOMO-HOMO, LUMO-LUMO, and HOMO-LUMO orbitals, thereby facilitating all possible excitations (Figures \ref{fig:ansatz} (b), \ref{fig:ansatz} (c) and \ref{fig:ansatz} (e)). Notably, ansatz 3 includes more RX and RY gates than the others, introducing additional parameters to the VQE, which may account for its superior accuracy. However, despite having fewer parameters, ansatz 5 outperforms ansatz 3, demonstrating that increasing the number of trainable parameters does not necessarily guarantee higher accuracy.

Two observations merit special attention: first, UCCSD, despite being applied across all active spaces, consistently failed to achieve chemical precision, with parameter counts of 26, 117, and 360 for the four-, six-, and eight-orbital spaces, respectively. Second, as expected, increasing the active space generally enhances the precision of the ansatzes (Table \ref{tab:energy}).

\begin{table}[H]
\centering
\caption{Energies obtained for each ansatz at a specified number of molecular orbitals (MOs) during the simulation of protocatechuic acid. The table displays the energy values achieved by the different ansätze, allowing for a direct comparison of their performance across the selected orbital configurations. Each cell represents the energy outcome for a specific ansatz, highlighting the variations in accuracy and efficiency depending on the approach used. This table provides insight into the effectiveness of each ansatz in approximating the reference ground state energy of the system.}
\begin{tabular}{l|lll|l}
 \cline{2-4}
& \multicolumn{3}{c|}{\textbf{Energy (Hartree)}}                                           &  \\ \cline{2-4}\multicolumn{1}{l|}{}         & \multicolumn{1}{c|}{\textbf{4 orbitals}} & \multicolumn{1}{c|}{\textbf{6 orbitals}} & \multicolumn{1}{c|}{\textbf{8 orbitals}} &  \\ \cline{1-4} \cline{1-4}
\multicolumn{1}{|l|}{\textbf{Ansatz 1}} & \multicolumn{1}{l|}{-545.830656}           & \multicolumn{1}{l|}{-546.346322}           & \multicolumn{1}{l|}{-546.280614}           &  \\ \cline{1-4} \cline{1-4}
\multicolumn{1}{|l|}{\textbf{Ansatz 2}} & \multicolumn{1}{l|}{-547.271782}           & \multicolumn{1}{l|}{-547.813201}           & \multicolumn{1}{l|}{-548.397301}           &  \\  \cline{1-4}
\multicolumn{1}{|l|}{\textbf{Ansatz 3}} & \multicolumn{1}{l|}{-548.099048}           & \multicolumn{1}{l|}{-548.754270}           & \multicolumn{1}{l|}{-549.760683}           &  \\   \cline{1-4}
\multicolumn{1}{|l|}{\textbf{Ansatz 4}} & \multicolumn{1}{l|}{-547.798908}           & \multicolumn{1}{l|}{-548.033164}           & \multicolumn{1}{l|}{-548.493309}           &  \\  \cline{1-4}
\multicolumn{1}{|l|}{\textbf{Ansatz 5}} & \multicolumn{1}{l|}{-548.099208}           & \multicolumn{1}{l|}{-548.754270}           & \multicolumn{1}{l|}{-549.759686}           &  \\  \cline{1-4}
\multicolumn{1}{|l|}{\textbf{Ansatz 6}} & \multicolumn{1}{l|}{-547.282408}           & \multicolumn{1}{l|}{-548.521130}           & \multicolumn{1}{l|}{-549.425126}           &  \\  \cline{1-4}
\multicolumn{1}{|l|}{\textbf{Reference}} & \multicolumn{1}{l|}{-548.099211}           & \multicolumn{1}{l|}{-548.756214}           & \multicolumn{1}{l|}{-549.928145}           &  \\ \cline{1-4}
\end{tabular}
\label{tab:energy}
\end{table}

\section{Conclusion} \label{sec:conclusion}

In the NISQ era, optimizing the Variational Quantum Eigensolver (VQE) is crucial for effectively determining the ground state energy of molecules. The choice of Ansatz and the selection of active spaces significantly influence VQE's performance, as these parameters directly affect the algorithm's accuracy and resource efficiency. Research indicates that a well-chosen Ansatz can enhance the convergence of VQE, allowing it to tackle more complex problems without excessive resource consumption. Additionally, the selection of active spaces must be carefully considered to balance computational feasibility with the fidelity of the results, as larger spaces can lead to increased computational demands. Furthermore, studies emphasize the importance of tailoring these choices to specific molecular systems to maximize VQE's efficacy, highlighting that a one-size-fits-all approach may not yield optimal results. Thus, strategic optimization of Ansätze and active spaces is essential for advancing quantum computational capabilities in the NISQ era.

In this study, we explored the application of the Variational Quantum Eigensolver (VQE) to model the ground state energies of molecular systems, focusing specifically on protocatechuic acid. By comparing the performance of six different ansatzes and systematically reducing the active space from eight to six and four molecular orbitals, we gained valuable insights into the impact of quantum circuit structure on simulation accuracy. Our findings indicate that ansatzes 3 and 5 consistently outperformed others, achieving chemical precision for smaller active spaces, while maintaining a balance between parameter count and computational efficiency.

Moreover, our results highlight the critical importance of ansatz selection when implementing VQE on near-term quantum devices. As we observed, the number of parameters does not necessarily correlate with increased accuracy, as evidenced by the superior performance of ansatz 5 despite its reduced complexity compared to ansatz 3. These findings underscore the need to optimize circuit design based on the specific requirements of the system under study, particularly in the context of limited qubit resources and noise-prone quantum hardware in the NISQ era.

Looking forward, expanding the active space and refining the ansatzes will be essential for extending VQE's applicability to larger and more complex molecules. Additionally, further research into noise-resilient quantum algorithms and hybrid quantum-classical approaches will be critical for improving accuracy and scalability. Our work contributes to this ongoing effort by demonstrating the practical viability of VQE for small molecular systems and laying the groundwork for future advancements in quantum computational chemistry.

\section{ACKNOWLEDGEMENTS}

This work has been partially supported by QuIIN - EMBRAPII CIMATEC Competence Center in Quantum Technologies, with financial resources from the PPI IoT/Manufatura 4.0 of the MCTI grant number 053/2023, signed with EMBRAPII. C. Cruz, G.F. de Jesus, L.Q. Galvão and M.H.F. da Silva thank the Bahia State Research Support Foundation (FAPESB) for financial support (grant numbers APP0041/2023 and PPP0006/2024). C. Cruz acknowledges Dr. Elias Brito for his invaluable contribution in shaping the initial formulation of this work.

\appendix

\section{Ansätze} 
\label{apend:ansatze}
In the literature, there are various ansatz proposals for tackling different problems, which can be summarized into two main classes: Physical Motivated Ansatz (PMA) and Hardware Heuristic Ansatz (HHA) \cite{cao2019quantum}. In the former, the ansatz is built taking into account the physical properties of each system to model a more accurate wave function, while in the latter, they are based on the architecture of quantum computers, aiming for greater performance efficiency \cite{cao2019quantum, romero2018strategies, mcardle2019variational, tang2021qubit, leone2022practical}. In the context of Quantum Chemistry simulations, certain ansatz of each class are highlighted, namely the Quantum Unitary Coupled Cluster (qUCC) ansatz from the PMA class and the Hardware Efficient (HE) ansatz from the HHA class.

\emph{Quantum Unitary Coupled Cluster ansatz}. This ansatz is based on the Coupled Cluster (CC) method, which is widely recognized as a prominent method in Quantum Chemistry simulations, given the possibility of obtaining relatively coherent results at a lower computational cost \cite{bartlett2007coupled, crawford2007introduction}. In short, the CC method consists of applying the excitation operator to the wave function, given a certain $\eta-Fock$ space, referring to the space of active orbitals \cite{romero2018strategies}. To do this, the excitation operator takes into account the contributions of the chosen excitations:

\begin{equation} \label{eq:excitation}
    T = \sum_{i=1}^{\eta} T_i
\end{equation}

where, for single and double excitations, in which case the method is called Coupled Cluster Singles and Doubles (CCSD), the operators are defined as:

\begin{equation} \label{eq:excitation-1}
    T_1 = \sum_{\substack{i \in occ \\ a \in virt}} t_a^i a_a^{\dagger}a_i
\end{equation}

\begin{equation} \label{eq:excitation-2}
    T_2 = \sum_{\substack{i > j \in occ \\ a > b \in virt}} t_{ab}^{ij} a_a^{\dagger}a_b^{\dagger} a_i a_j
\end{equation}

where \emph{occ} represents the occupied regions and \emph{virt} the unoccupied regions of the active space \cite{purvis1982full}. 
In this way, it is possible to represent the wave function as an exponential product of single and double excitation operators:
\begin{equation}
    \ket{CCSD} = e^{T} \ket{HF} = e^{T_1 + T_2}~. \ket{HF}
\end{equation}
where $\ket{HF}$ is the reference state associated with the Hartree-Fock state. Once the CC method formalism has been constructed, it is possible to obtain the energy of the Hamiltonian of this system from the Schr\" odinger equation:
\begin{equation} \label{eq: schger}
    e^{-T} H e^{T} \ket{HF} = E \ket{HF}~. 
\end{equation}
with $E$ being the energy of the ground state, defined as:
\begin{equation}
    E = \min_{\vec{\theta}} \bra{HF}e^{-T} H e^{T} \ket{HF}~.
\end{equation}

Considering the equations above, it is worth highlighting two properties of the operator $e^{-T} H e^{T}$, known as the similarity-transformed Hamiltonian: i) it can be expanded; ii) it is not unitary \cite{bartlett2007coupled, romero2018strategies, crawford2007introduction}. The first property can be verified from a Baker-Campbell-Hausdorff (BCH) expansion \cite{gilmore1974baker}, which results in a finite series:
\begin{multline}
e^{-T}He^T = H + [H, T] + \frac{1}{2}[[H, T], T] \\
+ \frac{1}{3!}[[[H, T], T], T] + \frac{1}{4!}[[[[H, T], T], T], T]
\end{multline}

Applied to equation (\ref{eq: schger}), we obtain a set of nonlinear equations with the energies and amplitudes of the system. Furthermore, an important aspect of the BCH expansion is that the system converges only under the assumption of a single reference state. As a result, the method generally performs poorly for strongly correlated systems.

In the second property of the similarity-transformed Hamiltonian, there is a limitation to the calculations, since the quantum operators must necessarily be unitary to respect the conservation of probabilities of the system \cite{Sakurai1993Modern}. As an alternative method, the Unitary coupled cluster (UCC) was proposed \cite{cao2019quantum}, known as UCCSD for single and double excitations:

\begin{equation}
    \ket{\psi} = e^{T - T^{\dagger}} \ket{HF}
\end{equation}

And the energy of the ground state:

\begin{equation}
    E = \min_{\vec{\theta}} \bra{HF}e^{T^{\dagger}-T} H e^{T - T^{\dagger}} \ket{HF}
\end{equation}

Note that the similarity-transformed Hamiltonian, $e^{T - T^{\dagger}}$, now corresponds to a unitary operator, since $T - T^{\dagger}$ is Hermitian. With this result, the BCH expansion results in an infinite series

\begin{multline}
e^{T^\dagger - T}He^T e^{-T^\dagger} = H + [H, T] + T^\dagger H \\
+ \frac{1}{2}([[H, T], T] + T^\dagger, T^\dagger, H + H, T, T^\dagger) + \ldots
\end{multline}

This implies that the UCC is an approximate method, regardless of the number of excitations chosen \cite{romero2018strategies}. Nevertheless, this approximation makes it possible to map the unitary operator onto a quantum circuit, an aspect that is impossible to achieve with the $e^{-T}$ operator, due to its non-unitarity. One possible mapping of these unitary operators onto quantum logic gates (qUCC) is the Trotterization \cite{kluber2023trotterization, babbush2015chemical}, which uses the Trotter-Suzuki formula: 

\begin{equation}
    U(\vec{t}) \approx U_{Trot}(\vec{t}) = \left ( \prod_j e^{\frac{t_j}{r} (T_j - T_j^{\dagger})}  \right )^{r} 
\end{equation}

In which $r$ is the Trotter number and $t_j$ is the amplitude of the coupled cluster. Considering the Jordan-Wigner transformations, it is possible to map the excitation operator into a product of Pauli matrices $P_k^j$:

\begin{equation}
    (T_j - T_j^{\dagger}) = i \sum_{k}^{2^{2l_j-1}} P_k^j
\end{equation}

Where $l_j$ is the degree of excitation of the $j$-th excitation operator $T_i$ \cite{romero2018strategies}. And finally, we obtain an operator that can be applied to a quantum circuit:

\begin{equation} \label{eq:qUCC-1}
     U(\vec{t}) \approx U_{Trot}(\vec{t}) = \left ( \prod_j e^{\frac{t_j}{r} (\sum_{k}^{2^{2l_j-1}} P_k^j)}  \right )^{r} 
\end{equation}

But since the excitation operator $(T_j - T_j^{\dagger})$ must commute with $P_j^k$, then we can rewrite eq. (\ref{eq:qUCC-1}) as:

\begin{equation} \label{eq:qUCC-2}
     U(\vec{t}) \approx U_{Trot}(\vec{t}) = \left ( \prod_j \prod_k^{2^{2l_j-1}} e^{ \left (i \frac{t_j}{r} P_j^k \right )} \right )^{r}
\end{equation}

The Pennylane library offers a series of commands that map the excitation operator described in eq. (\ref{eq:excitation}), considering single (\ref{eq:excitation-1}) and double excitations (\ref{eq:excitation-2}) in order to construct the UCCSD ansatz. A useful method in the library is \emph{AllSinglesDoubles()}, in which, given the optimization parameters and the active space ($\eta-Fock$), it indicates the gates and parameters needed to construct the parameterized unitary operator of eq. (\ref{eq:qUCC-2}). Using Jordan Wigner transformation, the number of gates scales as $O (N^3 \eta^2)$, while the number of parameters scales as $\left ( \frac{N - \eta}{2}\right ) \left ( \frac{\eta}{2}\right ) + (N - \eta)(\eta) < O(N^2 \eta^2)$, where N is the number of spin orbitals and $\eta$ is the number of electrons \cite{romero2018strategies}.

\emph{Hardware Efficient ansatz}. In contrast to qUCC, the HE ansatz does not consider any properties of quantum phenomena in its construction. This was one of the first ansatz proposed for the use of VQE, given its versatility and compatibility with the architecture of some current quantum computers \cite{cao2019quantum}. In short, the ansatz seeks to perform rotations considering the most relevant qubits for the problem, as well as the connectivity associated with its architecture \cite{leone2022practical}. In this regard, it is assumed that a HE with more rotations implies a more accurate result, although it is computationally more expensive. 

From the point of view of simulating molecules in quantum circuits, the HE ansatz is usually less accurate, since it disregards the couplings of the electrons in the molecule to the detriment of the connectivity of the qubits \cite{leone2022practical}. Despite this, some studies have proposed changes to the HE ansatz circuits to adapt to the structure of the system and obtain better results \cite{kandala2017hardware, mccaskey2019quantum, zheng2023sncqa}.

\bibliographystyle{unsrt}
\bibliography{references}

\end{document}